\documentclass[prl,twocolumn]{revtex4}
\usepackage{amssymb}
\usepackage{graphicx}
\usepackage[latin1]{inputenc}
\begin{document}
\title{Electromotive force and internal resistance of an electron pump}
\author{Miguel Rey}
 \email{miguel.rey@uam.es}
 \affiliation{Departamento de Física Teórica de la Materia Condensada and
  Instituto Nicolás Cabrera,
 Universidad Autónoma de Madrid, E-28049 Madrid, Spain}
\author{Fernando Sols}
 \email{fernando.sols@uam.es}
 \affiliation {Departamento de Física Teórica de la Materia Condensada and
  Instituto Nicolás Cabrera,
 Universidad Autónoma de Madrid, E-28049 Madrid, Spain}
\date{\today}
\begin{abstract}
We present a scattering theory of the electromotive force and
internal resistance of an electron pump. The characterization of
the device performance in terms of only two parameters requires
the assumption of incoherent multiple scattering within the
circuit and complete thermalization among electrons moving in a
given direction. The electromotive force is shown to be of the
order of the driving frequency in natural units. In an open setup,
the electromotive force adds to the voltage difference between
reservoirs to drive the current, both facing a contact resistance
which is absent in the case of a closed circuit of uniform width.
\end{abstract}
\maketitle
Electrons pumps favor electron motion in a given direction by
combining nonlinear ac driving and some asymmetry in the spatial
structure or in the temporal signal. This rectification of
electron motion generates current in the absence of a net dc
voltage bias
\cite{swit99,thou83,staf96,pede98,brou98,wagn99,kim02,lehm03}.
Most theoretical calculations have dealt so far with the
calculation of the {\it pump current}, which is the current
flowing through the device when the electrons incident from both
sides are characterized by the same chemical potential. In
practice, however, one is likely to be interested in the
performance of the pump as a circuit component, something which
cannot be predicted from the mere knowledge of the pump current.
This creates the need to characterize the electron pump as a
battery with a certain {\it electromotive force} and {\it internal
resistance}. Although the electromotive force can in principle be
obtained from the dc bias that exactly cancels the pump current
\cite{swit99,wagn99}, its derivation within a unified and general
scheme seems desirable. On the other hand, there is no obvious
ansatz for the calculation of the internal resistance. A potential
application of this new class of devices is the generation of
current in small closed circuits not attached to broad wires
acting as electron reservoirs. Such a setup is schematically
depicted in Fig. 1. One may also consider a pump in series with a
resistor, both within a lead that couples to large reservoirs
through ideal contacts, as indicated in Fig. 2.
In this article, we derive formulae which express the battery
parameters in terms of the transmission and reflection
probabilities for electrons crossing the pump and the resistor.
The scattering theory here presented attempts to play a role for
the electromotive force and internal resistance of an electron
pump similar to that which the Landauer-B\"uttiker theory has
represented for the conductance of nanostructures \cite{land57}.
An important difference, however, is that the scenario which we
investigate requires a more coarse-grained description if we wish
to uniquely characterize the circuit performance of the pump in
terms of a small set of parameters. Such an effective
self-averaging of the device performance requires the assumption
of electron decoherence between the pump and the resistor in
series. Unlike in Ref. \cite{butt86a}, we assume that effective
phase randomization can be achieved with negligible
backscattering. This hypothesis is supported by the fact that a
minor distortion of the environment suffices to induce electron
dephasing, while a more continued interaction is needed to change
the electron energy or direction appreciably \cite{ster90}. We
find that reflectionless decoherence between the circuit elements
is still insufficient to permit the characterization in terms of
only {\it two} parameters, and not twice as many as available
transverse channels. To achieve a simple and manageable
description we must assume that, within the leads, electrons
moving in each direction are characterized by a single chemical
potential. The adequacy of this assumption, or its replacement by
a weaker one within a model of comparable tractability, deserves
further study.
The total current through a two-lead multimode structure in the
presence of local ac driving may be written
\begin{equation}
I = \frac{e}{h}\int dE_i [ f(E_i-\mu^{\rm{in}}_R)T_{LR}(E_i) -
f(E_i-\mu^{\rm{in}}_L)T_{RL}(E_i)]
\end{equation}
\begin{equation}
T_{LR}(E_i) \equiv \sum_{a \in L} \sum_{b \in R} \int dE_f
T^{LR}_{ab}(E_f, E_i),
\end{equation}
$T^{LR}_{ab}(E_f, E_i)$ being the probability distribution that an
electron incident from the right lead in channel $b$ with energy
$E_i$ is transmitted into channel $a$ of the left lead with energy
$E_f$. For future convenience, we assume $I>0$ when current flows
from right to left. The chemical potentials
$\mu^{\rm{in}}_L,\mu^{\rm{in}}_R$ characterize the population of
incoming electrons.
The pump effect is based on the existence of an asymmetry between
the left-to-right and right-to-left transmissions. Thus it is
convenient to define:
\begin{eqnarray} \label{average}
T(E_i) &\equiv& [T_{LR}(E_i) + T_{RL}(E_i)]/2 \\
\label{difference} \delta T(E_i) &\equiv& T_{LR}(E_i) -
T_{RL}(E_i)\ .
\end{eqnarray}
If we linearize $f(E-\mu^{\rm in}_{L,R})$ around a common
reference chemical potential $\mu_0$, we may write the total
current as the sum of a bias and a pump contribution
\begin{eqnarray}
I&=&I_B+I_P \\
I_B &\equiv& \frac{e}{h} \Delta\mu^{\rm in}
 \int dE_i [-f'(E_i-\mu_0)] T(E_i) \\
\label{I-pump-general} I_P &\equiv& \frac{e}{h} \int f(E_i-\mu_0)
\delta T(E_i) \ ,
\end{eqnarray}
with $\Delta\mu^{\rm in} \equiv \mu^{\rm{in}}_R- \mu^{\rm{in}}_L$.
Hereafter, we take $\mu_0 \equiv 0$, although we note that (unlike
$I_B$) $I_P$ does depend on $\mu_0$.
\begin{figure}[t]
\includegraphics[width=8cm]{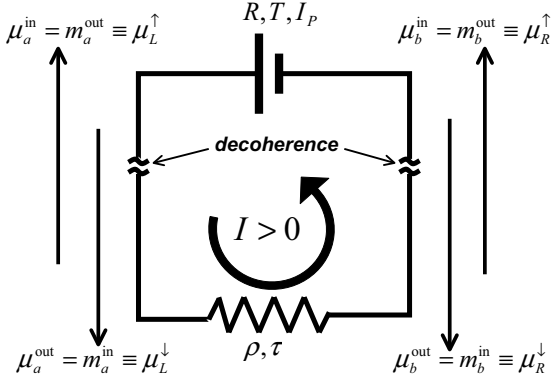}
\caption{Schematic representation of an electron pump in series
with a generic resistor within a closed circuit where current
flows thanks to the action of the pump}
\end{figure}
Let us focus on the current flow in a given channel $a$ on e.g.
lead $L$.  For convenience, we define $\tilde{I}_{\alpha}\equiv
(h/e)I_{\alpha}$ for all future current contributions. To achieve
a better perspective, we momentarily abandon the assumption that
the chemical potential is channel independent. The total current
through channel $a\in L$ can then be written:
\begin{equation}
\tilde{I}_{a}= -\mu^{\rm{in}}_a + \mu^{\rm out}_a,
\end{equation}
where $\mu_a^{\rm in(out)}$ characterizes the population of
electrons in $L$ approaching (leaving) the pump. We note that,
even if the ``in'' population is rigorously thermal, the ``out''
population is not. However, one can always find a suitably defined
chemical chemical potential $\mu_a^{\rm out}$ that reproduces the
same current flow (and, in one dimension, the same current density
\cite{sols99}). Like the total current, this {\it outgoing
chemical potential} has a ``bias'' and a ``pump'' contribution,
\begin{eqnarray}
\mu^{\rm out}_a &=& \mu^{\rm out,B}_a+ \mu^{\rm out,P}_a \\
\mu^{\rm out,B}_a &=& \sum_b S_{ab}\mu^{\rm{in}}_b \\
\mu^{\rm out,P}_a &=& \tilde{I}_{P,a} \ ,
\end{eqnarray}
where $\tilde{I}_{P,a}$ is the pump current in channel $a$
($\sum_a \tilde{I}_{P,a}=\tilde{I}_P$). Since the bias
contribution depends only on the symmetrized probability [see Eq.
(\ref{average})], we have $S_{ab}=S_{ba}$. On the other hand,
unitarity requires $\sum_b S_{ab} = 1$. The term $\mu_a^{\rm
out,P}$ accounts for the excess (or defect) of electrons generated
by the pump. It reflects the fact that an operating battery
creates a {\it population imbalance} which ultimately drives the
current through the circuit.
Assume that, in series with the pump, we introduce a resistor
which is also characterized in terms of its scattering
probabilities. The resulting circuit is schematically depicted in
Fig. 1. Being the resistor a passive element, its flow equations
do not include a pump term. We write
\begin{equation} \label{resistor}
\tilde{I}_a = m^{\rm{in}}_a - \sum_b \sigma_{ab} m^{\rm{in}}_b
\equiv m^{\rm{in}}_a - m^{\rm{out}}_a,
\end{equation}
where $m^{\rm{in(out)}}_a$ is the chemical potential for the
electrons approaching (leaving) the resistor, and
$\{\sigma_{ab}\}$ are the scattering probabilities by the
resistor, which obey $\sigma_{ab}=\sigma_{ba}$ and
$\sum_a\sigma_{ab}=1$. The sign convention in Eq. (\ref{resistor})
is different from that used in the pump equations because
counterclockwise current is taken to be positive (see Fig. 1). Now
we note that the ``out'' population of the pump is the ``in''
population of the resistor, and viceversa. We seal this
equivalence by establishing a common notation. For $a \in L$ we
write
\begin{eqnarray}
\mu^{\rm{in}}_a = m^{\rm{out}}_a \equiv \mu^{\uparrow}_{L,a} \\
\mu^{\rm{out}}_a = m^{\rm{in}}_a \equiv \mu^{\downarrow}_{L,a}\ ,
\end{eqnarray}
and similarly for $a \in R$. The vertical arrows refer to the
direction of movement within the convention of Fig. 1.
\begin{figure}[t]
\includegraphics[width=8cm]{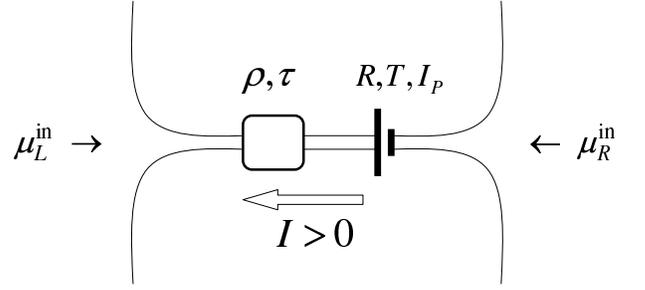}
\caption{Electron pump in series with a scattering barrier, both
within a multichannel wire that couples through ideal contacts to
large electron reservoirs. Both the action of the pump and the
difference between the reservoir chemical potentials contribute to
drive the current through the wire}
\end{figure}
Although a solution of the flow equations that would permit us to
predict the total current in terms of
$\{I_{P,a},S_{ab},\sigma_{ab}\}$ is formally possible, our real
goal is the characterization of the pump in terms of two
parameters. To achieve this objective, we have to introduce the
simplifying assumption that electrons flowing in a given direction
within a lead are all characterized by the same chemical
potential. We express it as
\begin{equation}
\mu^{\uparrow \downarrow}_{L,a} = \mu^{\uparrow \downarrow}_L
\hspace{0.5cm} \mu^{\uparrow \downarrow}_{R,a} = \mu^{\uparrow
\downarrow}_R \hspace{0.5cm} \forall a \in L,R\ .
\label{equiv-chem-pot}
\end{equation}
Hereafter we differentiate between reflection and transmission
probabilities: $ S_{ab} \rightarrow R_{ab}, T_{ab}$ and
$\sigma_{ab} \rightarrow \rho_{ab}, \tau_{ab}$. We introduce the
notation
\begin{eqnarray} \label{RaTa}
R_a \equiv \sum_b R_{ab} \hspace{0.5cm} T_a \equiv \sum_b T_{ab}
\hspace{0.5cm} (R_a + T_a = 1)\ , \\
\label{RT} R \equiv \sum_a R_a
\hspace{0.5cm} T \equiv \sum_a T_a \hspace{0.5cm} (R + T = N)\ ,
\end{eqnarray}
$N$ being the number of transverse channels. The resistor
parameters $\rho_a,\tau_a,\rho,\tau$ are defined analogously. We
introduce an average ``pump chemical potential'' $\mu_P \equiv
\sum_a \mu^{\rm out, P}_{a}/N=\tilde{I}_P/N$.
The total current can now be written
\begin{equation} \label{total-I}
\tilde{I} = N(\mu^{\downarrow}_L - \mu^{\uparrow}_L) =
N(\mu^{\uparrow}_R - \mu^{\downarrow}_R)\ .
\end{equation}
These four chemical potential are not independent but are rather
related by the flow equations
\begin{eqnarray} \label{L-down}
N\mu^{\downarrow}_L &=& R \mu^{\uparrow}_L + T \mu^{\uparrow}_R + N \mu_P \\
\label{L-up} N\mu^{\uparrow}_L &=& \rho \mu^{\downarrow}_L + \tau
\mu^{\downarrow}_R \\
\label{R-down}
N\mu^{\downarrow}_R &=& T \mu^{\uparrow}_L + R \mu^{\uparrow}_R - N \mu_P \\
\label{R-up} N\mu^{\uparrow}_R &=& \tau \mu^{\downarrow}_L + \rho
\mu^{\downarrow}_R  \ ,
\end{eqnarray}
which are physically transparent. The different sign carried by
the pump contribution $N\mu_{P}$ in Eqs. (\ref{L-down}) and
(\ref{R-down}) expresses the fact that, when $\mu_P>0$, there is
an excess of outgoing electrons on the left of the pump and a
corresponding defect of outgoing electrons on the right. When Eqs.
(\ref{L-down}) -- (\ref{R-up}) are introduced into Eq.
(\ref{total-I}), we obtain for the total current
\begin{eqnarray} \label{I-T-up}
\tilde{I} &=& -T \Delta \mu^{\uparrow} + N \mu_P \\
\label{I-tau-down} &=& \tau \Delta \mu^{\downarrow}\ ,
\end{eqnarray}
where the chemical potential differences $\Delta \mu^{\uparrow
\downarrow} \equiv \mu^{\uparrow \downarrow}_L - \mu^{\uparrow
\downarrow}_R$ satisfy the relations
\begin{eqnarray} \label{dif-down}
N \Delta \mu^{\downarrow} &=& (R-T) \Delta \mu^{\uparrow} + 2
N\mu_P
\\ \label{dif-up}
N \Delta \mu^{\uparrow} &=& (\rho-\tau) \Delta \mu^{\downarrow} \
..
\end{eqnarray}
We may solve for $\Delta \mu^{\uparrow \downarrow}$ in Eqs.
(\ref{dif-down}) and (\ref{dif-up}) and introduce the solutions in
either (\ref{I-T-up}) or (\ref{I-tau-down}) to obtain
\begin{equation} \label{I-scattering}
\tilde{I} = \frac{ (N/T)I_P } {
\rho/\tau+ R/T } \ .
\end{equation}
Calculating the electromotive force ${\cal V}_{\rm emf}$ and the
internal resistance ${\cal R}_{i}$ amounts to finding a relation
\begin{equation} \label{I-circuit}
I = \frac{ {\cal V}_{\rm{emf}}}{{\cal R} + {\cal R}_{i}}\ ,
\end{equation}
where ${\cal R}$ is a suitably defined resistance for the
resistor. Comparison of Eqs. (\ref{I-scattering}) and
(\ref{I-circuit}) uniquely leads to the result
\begin{eqnarray} \label{final-V-emf}
{\cal V}_{\rm{emf}} &=& \frac{h}{e^2} \frac{I_P}{T}
\\
\label{final-R-i} {\cal R}_{i} &=& \frac{h}{Ne^2} \frac{R}{T}
\ ,
\end{eqnarray}
provided that
\begin{equation} \label{final-R}
{\cal R} = \frac{h}{Ne^2} \frac{\rho}{\tau}
\ .
\end{equation}
The prefactors have been chosen to make ${\cal V}_{\rm emf}$ an
intensive quantity while ${\cal R} ,{\cal R}_{i}\sim N^{-1}$ as $N
\rightarrow \infty$.
We may apply our results for ${\cal V}_{\rm emf}$ and ${\cal R}_i$
to the analytically  solvable pipeline model, which assumes that
transmission takes place only within a single pair of channels
\cite{wagn99}. It can be expressed as:
\begin{equation}
T_{ab}^{LR}(E_f, E_i) = J\delta_{ab}\delta(E^z_f- E_2)
\delta(E^z_i- E_1) \ .
\end{equation}
Here $E_{\alpha}^z$ ($\alpha=i,f$) is the energy in the direction
perpendicular to the planar structure and $(E_2-E_1)/\hbar=\omega
>0 $ is the driving frequency. The other scattering probabilities
are determined by time-reversal symmetry in the presence of
coherent ac driving [$T_{ab}^{LR}(E, E')=T_{ba}^{RL}(E',E)$] and
unitarity. For three dimensions, the single pipeline model yields
\begin{equation}
T = DJ
\end{equation}
where $D = Am/2\pi\hbar^2$ is the two-dimensional transverse
density of states, $A$ being the interface area. Preservation of
unitarity requires $DJ<N$. The total pump current is
\begin{equation}
 I_P = eDJ\omega/2\pi \ ,
\end{equation}
so we interpret $eDJ$ as the pumped charge per cycle. For the
circuit parameters we obtain
\begin{eqnarray} \label{emf-pump}
{\cal V}_{\rm{emf}} &=& \hbar \omega/e
\\ \label{int-resist-pump}
{\cal R}_{i} &=& \frac{h}{Ne^2} \frac{N-DJ}{DJ}
\end{eqnarray}
The result that the electromotive force is just $\hbar \omega/e$,
independently of the transmittivity  $J$, is remarkable if one
looks at the general structure of Eqs. (\ref{I-pump-general}),
(\ref{RaTa}), (\ref{RT}), and (\ref{final-V-emf}), but could have
been expected from the notion that the pipeline model allows only
for an energy gain $\hbar \omega$ as the electron is pumped from
right to left, regardless of the total electron flow. We readily
conclude that, in a more general pump structure, ${\cal
V}_{\rm{emf}} \sim \hbar \omega/e$, in agreement with Refs.
\cite{swit99,wagn99}. By contrast, the internal resistance is very
sensitive to the transmittivity of the pump. In particular, we
note that ${\cal R}_{i}\rightarrow \infty$ as $J\rightarrow 0$.
We now turn our attention to an open setup where the pump and
resistor stay in series within a lead coupled through ideal
contacts to broad electron reservoirs. As indicated in Fig. 2, the
chemical potentials in the reservoirs characterize the population
of the incoming electrons. Hence, we refer to them also as
$\mu_{L,R}^{\rm in}$. One may perform an analysis similar to that
described for the closed geometry of Fig. 1. After some algebra,
one obtains
\begin{equation} \label{I-open} I=(e/h)T'(
e{\cal V}_{\rm{emf}}+\Delta\mu^{\rm in}) \ ,
\end{equation}
where
${\cal V}_{\rm{emf}}$ is given by Eq. (\ref{final-V-emf}), and
$T'\equiv (T\tau/N)/(1-R\rho/N^2)$ is the the average transmission
through the compound structure formed by the pump plus the
resistor.
Interestingly, Eq. (\ref{I-open}) can also be written as
\begin{equation} \label{I-open-contact}
I=\frac{{\cal V}_{\rm emf}+\Delta \mu^{\rm in}/e}{ (h/Ne^2)+{\cal
R}+{\cal R}_{i} } \ ,
\end{equation}
where the resistances ${\cal R}$ and ${\cal R}_i$ are given by
Eqs. (\ref{final-R}) and  (\ref{final-R-i}) respectively. Thus we
see that, within an open geometry, the pump electromotive force
adds to the voltage bias generated by the potential difference
between the two electron reservoirs. This confirms the intuitive
expectation that ${\cal V}_{\rm{emf}}$ can be obtained from the
the voltage difference $\Delta\mu^{\rm in}$ needed to cancel the
pump current \cite{swit99,wagn99}.
A striking difference between Eqs. (\ref{I-circuit}) and
(\ref{I-open-contact}) is the role of the contact resistance
$h/Ne^2$, which is absent in the case of a closed structure.
Comparison of the underlying models suggests that the contact
resistance disappears under the assumption that the flow of
outgoing electrons on the left of the resistor-pump structure of
Fig. 2 is identified with the the flow of incoming electrons from
the right, and equivalently for electrons moving in the opposite
direction. We conclude that, within a closed circuit of uniform
width, there is no natural lower limit to the resistance that the
electron current must face as it is generated by the electron
pump. This result appears reasonable if one notes that contact
resistances along the circuit are generated by narrow-wide
contacts where the width of the wire (and thus the number of
available electron channels) changes \cite{imry86}.
The denominators of Eqs. (\ref{I-circuit}) and
(\ref{I-open-contact}) suggest that the resistances which we have
introduced should be additive. Unfortunately, the ratios $R/T$ and
$\rho/\tau$ cannot be guaranteed to be additive in general. That
is possible only in one dimension, where the ration $R/T$ is known
to be additive for barriers compounded incoherently \cite{datt95},
or, for multichannel wires, in the particular case where the
scattering probabilities are independent of the channel index
($R_{ab}=R/N^2$, $T_{ab}=T/N^2$, and similarly for
$\rho_{ab},\tau_{ab}$). In such a case, the assumption of a common
chemical potential for all incoming electrons ($\mu_{a}^{\rm
in}=\mu_{L,R}^{\rm in}$ for all $a \in L,R $) automatically
guarantees an outgoing population with a common chemical for all
outgoing electrons ($\mu_{b}^{\rm out}=\mu_{L,R}^{\rm out}$ for
all $b \in L,R$). Within that scheme, the assumption of a common
chemical potential for all electrons moving in a given direction
[see Eq. (\ref{equiv-chem-pot})] is internally consistent in the
sense that a scenario may be conceived where the outgoing
population from a barrier or pump is guaranteed to be a suitable
incoming population for the following obstacle. Interestingly, it
is also in the channel-independent scattering limit where the
resistance defined in Ref. \cite{butt85} becomes additive and
equivalent to the resistance defined in Eq. (\ref{final-R}).
On closer inspection, one realizes that the assumption of
channel-independent scattering is hard to justify within an
independent electron picture, where no naturally additive
resistance can be defined for multichannel wires without invoking
impurity averaging. In particular, such a hypothesis is not
satisfied by the pipeline model invoked above, since its
transmission depends on the perpendicular energy. We conclude that
the question of the definition of an electron pump internal
resistance is directly connected to the discussion on the
additivity of resistances in multichannel wires. As long as this
fundamental issue is not satisfactorily resolved, the transport
equations (\ref{I-circuit}) and (\ref{I-open-contact}) [as
complemented by (\ref{final-V-emf}), (\ref{final-R-i}), and
(\ref{final-R})] which we have derived will have to be viewed as
approximations obtained from a reasonable and appealing scheme.
This research has been supported by the MCyT (Spain) under Grant
No. BFM2001-0172, the EU RTN Programme under Contract No.
HPRN-CT-2000-00144, and the Ram\'on Areces Foundation.

\end{document}